\documentclass[11pt]{article}
\usepackage{amssymb, amsmath, amsthm, hyperref, tikz}
\usepackage[left=1in,top=1in,right=1in]{geometry}
\date{}
\allowdisplaybreaks

\theoremstyle{definition}
\newtheorem{theorem}{Theorem}[section]
\newtheorem{lemma}[theorem]{Lemma}

\newtheorem{claim}[theorem]{Claim}

\counterwithin{equation}{section}

\title{On balanced circuits in uniform rank-three oriented matroids}
\author{Ji Zeng\thanks{Alfréd Rényi Institute of Mathematics, Budapest 1053, Hungary. Supported by ERC Advanced Grants ``GeoScape'', no. 882971 and ``ERMiD'', no. 101054936. Email: {\tt zeng.ji@renyi.hu}.}}

\date{}

\begin{document}

\maketitle

\begin{abstract}
A set of four points $\{p_1,p_2,p_3,p_4\}$ on the sphere is a balanced quadruple if there are four real numbers $s_1,s_2,s_3,s_4$, two positive and two negative, such that $s_1p_1+s_2p_2+s_3p_3+s_4p_4 = 0$. Streltsova and Wagner proved that $n$ points on the sphere determine at least
\[ \frac{1}{4} \left\lfloor \text{\makebox[0pt][l]{$\dfrac{n}{2}$}\phantom{$\frac{1}{2}$}} \right\rfloor \left\lfloor\frac{n-1}{2}\right\rfloor \left\lfloor\frac{n-2}{2}\right\rfloor \left\lfloor\frac{n-3}{2}\right\rfloor\]
many balanced quadruples provided any three points are linearly independent. We extend this result from spherical point configurations to rank-three oriented matroids.
\end{abstract}

\section{Introduction}

Let $P$ be a configuration of $n$ points on the sphere $\mathbb{S}^2$ in \textit{general position}, meaning that any three corresponding vectors in $\mathbb{R}^3$ are linearly independent. Every four-element subset of $P$ then has a unique linear dependence up to multiplication by a nonzero scalar. We call a four-element subset a \textit{balanced quadruple} if its linear dependence has exactly two positive and two negative scalars. Streltsova and Wagner~\cite{SW} proved that $P$ has at least $H(n)$ balanced quadruples, where
\begin{equation*}
H(n) = \frac{1}{4} \left\lfloor \text{\makebox[0pt][l]{$\dfrac{n}{2}$}\phantom{$\frac{1}{2}$}} \right\rfloor \left\lfloor\frac{n-1}{2}\right\rfloor \left\lfloor\frac{n-2}{2}\right\rfloor \left\lfloor\frac{n-3}{2}\right\rfloor.
\end{equation*}

An \textit{oriented matroid} is a combinatorial generalization of point configurations. A rank-three oriented matroid $M$ consists of a set $E$, called \textit{elements}, and a non-trivial alternating function $\chi:E^3\to\{-1,0,+1\}$, called \textit{chirotope}, such that for any $a_1,a_2,a_3,b_1,b_2,b_3\in E$, \begin{equation*}
    \begin{cases}
        \chi(b_1,a_2,a_3)\chi(a_1,b_2,b_3) \geq 0\\
        \chi(b_2,a_2,a_3)\chi(b_1,a_1,b_3) \geq 0\\
        \chi(b_3,a_2,a_3)\chi(b_1,b_2,a_1) \geq 0
    \end{cases} \implies~ \chi(a_1,a_2,a_3)\chi(b_1,b_2,b_3) \geq 0.
\end{equation*} $M$ is said to be \textit{uniform} provided $|\{a,b,c\}|=3$ implies $\chi(a,b,c) \neq 0$. A point configuration $P\subset \mathbb{S}^2$ gives an oriented matroid $M$ through
\begin{equation*}
    \chi(a,b,c)=\text{sign}\det(a,b,c) \quad \forall a,b,c \in P.
\end{equation*} Clearly, $P$ being in general position is equivalent to $M$ being uniform.

In basic linear algebra, four points $p_1,p_2,p_3,p_4$ on $\mathbb{S}^2$ satisfy
\begin{equation*}
    \det(p_2,p_3,p_4) \cdot p_1- \det(p_1,p_3,p_4) \cdot p_2+ \det(p_1,p_2,p_4) \cdot p_3- \det(p_1,p_2,p_3) \cdot p_4=0.
\end{equation*} Motivated by this, one can define \textit{circuit} $X = (X_a, X_b, X_c, X_d)$ on four elements $a,b,c,d$ by \begin{equation*}
    X_a = \chi(b,c,d),\quad X_b = -\chi(a,c,d),\quad X_c = \chi(a,b,d),\quad X_d = -\chi(a,b,c).
\end{equation*} The \textit{type} of the quadruple $\{a,b,c,d\}$ is defined as the smaller number between the plus-count and the minus-count in $X$. For example, $(+,+,+,+)$ means type 0, $(+,-,-,-)$ means type 1, and $(+,-,+,-)$ means type 2. The computation of circuit $X$ depends on an order $a \prec b \prec c \prec d$ we implicitly give. However, by the alternating property of $\chi$, type is independent of this order hence well-defined. By our observation above from basic linear algebra, a balanced quadruple on the sphere corresponds to a quadruple of type 2 in the language of oriented matroids. For this reason, we use balanced quadruples to call quadruples of type 2.

\begin{theorem}\label{main}
Every uniform rank-three oriented matroid on $n$ elements determines at least $H(n)$ balanced quadruples.
\end{theorem}

The bound by Streltsova and Wagner~\cite{SW} is sharp where the equality holds for the so called coneighborly configurations. In fact, their result (as one of the many nice results in~\cite{SW}) confirms the spherical drawing case of Hill's conjecture that any drawing of the complete graph of order $n$ has at least $H(n)$ crossings. And it is known that Hill's conjecture, if true, would be a sharp bound. Our result can be seen as evidence in favor of this conjecture.

\section{Preliminaries}

A \textit{pseudocircle} on $\mathbb{S}^2$ is the image of a great circle under a self-homeomorphism of $\mathbb{S}^2$; its complement consists of two open regions called its \textit{sides}. It is \textit{centrally symmetric} if it is invariant under the antipodal map and the antipodal map exchanges its two sides. A collection of pseudocircles is called an \textit{arrangement} if any two of them intersect exactly twice and cross properly at their intersections.


\begin{lemma}\label{central}
    For a uniform rank-three oriented matroid $M = (E,\chi)$, there exist points $p_e \in \mathbb{S}^2$ for all $e\in E$, and centrally symmetric pseudocircles $C_{ab}$ for all $\{a,b\} \in \binom{E}{2}$, satisfying the following conditions:\begin{enumerate}
        \item all $2n$ points $\{p_e,-p_e\}_{e\in E}$ are distinct;
        \item all pseudocircles $\{C_{ab}\}_{\{a,b\} \in \binom{E}{2}}$ form an arrangement;
        \item the pseudocircle $C_{ab}$ contains exactly $p_a$ and $p_b$ from $\{p_e\}_{e\in E}$;
        \item for $\{a,b,c,d\} \in \binom{E}{4}$, $\chi(a,b,c) = \chi(a,b,d)$ if and only if $p_c$ and $p_d$ are on the same side of $C_{ab}$.
    \end{enumerate}
\end{lemma}

Without the central symmetry condition, this is a standard fact known as TYPE II representation (see Theorem~5.3.6 and Proposition~6.3.6 in \cite{BLSWZ}). Given a TYPE II representation of $M$, there exists a homeomorphism of the sphere that transforms its pseudocircles into centrally symmetric ones (see Theorems~5.2.1 and~5.1.6 in \cite{BLSWZ}).

A \textit{pseudoline} in the plane $\mathbb{R}^2$ is the image of a straight line under a self-homeomorphism of $\mathbb{R}^2$. In a pseudoline arrangement, every two pseudolines have exactly one intersection and cross properly there. A \textit{generalized configuration} in $\mathbb{R}^2$ consists of labelled points $q_e$ and such an arrangement containing, for each pair $\{a,b\}$, a pseudoline $L_{ab}$ through $q_a$ and $q_b$, with no third labelled point on it. An \textit{allowable sequence} on a set $E$ of labels is a sequence $\pi_0,\dots,\pi_T$ of linear orders on $E$ in which consecutive orders differ by one swap of adjacent entries, every pair $\{a,b\}\in \binom{E}{2}$ is swapped exactly once, and $\pi_T$ is the reverse of $\pi_0$. Necessarily $T=\binom{|E|}{2}$. For a swap $\pi_{t-1} \to \pi_t$ involving two labels $a$ and $b$, the labels preceding $a,b$ are called \textit{prefix}, and the labels succeeding $a,b$ are called \textit{suffix}, of this swap.

\begin{lemma}\label{allowable}
    Given a generalized configuration $\{q_e\}$ for $e\in E$ and $L_{ab}$ for $\{a,b\} \in \binom{E}{2}$, there exists an allowable sequence on $E$ such that for every $\{a,b\}\in \binom{E}{2}$, the points labelled by the prefix and the suffix of the swap involving $a$ and $b$ are on opposite sides of $L_{ab}$.
\end{lemma}

\begin{figure}[t]
\centering
\definecolor{wireone}{RGB}{237,28,44}
\definecolor{wiretwo}{RGB}{0,157,215}
\definecolor{wirethree}{RGB}{88,47,153}
\definecolor{wirefour}{RGB}{0,145,72}
\begin{tikzpicture}[
  x=1.25cm,
  y=.72cm,
  wire/.style={line width=1.25pt,line cap=round,line join=round,
               rounded corners=2pt},
  label/.style={font=\large}
]
  \draw[wire,draw=wireone]
    (0,0)--(2.05,0)--(2.55,1)--(3.25,1)--(3.75,2)
    --(4.25,2)--(4.75,3)--(8,3);
  \draw[wire,draw=wiretwo]
    (0,1)--(2.05,1)--(2.55,0)--(5.45,0)--(5.95,1)
    --(6.45,1)--(6.95,2)--(8,2);
  \draw[wire,draw=wirethree]
    (0,2)--(1.05,2)--(1.55,3)--(4.25,3)--(4.75,2)
    --(6.45,2)--(6.95,1)--(8,1);
  \draw[wire,draw=wirefour]
    (0,3)--(1.05,3)--(1.55,2)--(3.25,2)--(3.75,1)
    --(5.45,1)--(5.95,0)--(8,0);

  \node[label,left=5pt]  at (0,3) {$4$};
  \node[label,left=5pt]  at (0,2) {$3$};
  \node[label,left=5pt]  at (0,1) {$2$};
  \node[label,left=5pt]  at (0,0) {$1$};
  \node[label,right=5pt] at (8,3) {$1$};
  \node[label,right=5pt] at (8,2) {$2$};
  \node[label,right=5pt] at (8,1) {$3$};
  \node[label,right=5pt] at (8,0) {$4$};
\end{tikzpicture}
\caption{A wiring diagram which gives an allowable sequence $1234 \to 1243 \to 2143 \to 2413 \to 2431 \to 4231 \to 4321$.}
\label{wiring}
\end{figure}
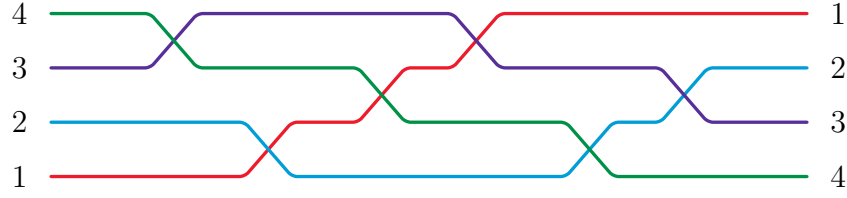

To see this, one can apply point-pseudoline duality to the given generalized configuration while preserving the above-below relationship, and transform the dual configuration into a wiring diagram (see Theorems~5.1.6 and~5.1.4 in \cite{FG}). Then one can read the swaps of the wanted allowable sequences as the crossings in this wiring diagram from left to right; see Figure~\ref{wiring}.

\section{Proof}

\subsection{Reduction to odd order}

Throughout this proof, we assume that $n$ is an odd integer and write $n = 2m + 1$. This does not affect the generality of our proof due to the following claim.

\begin{claim}
    If Theorem~\ref{main} holds for an odd integer $n$, then Theorem~\ref{main} holds for $n+1$ as well.
\end{claim}

Indeed, suppose $M' = (E',\chi')$ is a uniform rank-three oriented matroid with $|E'|=n+1$, we can apply Theorem~\ref{main} to the oriented matroid $M$ induced on $E' \setminus e$ for each $e\in E'$. Every application of Theorem~\ref{main} produces $H(n)$ balanced quadruples in $M'$. On the other hand, a fixed balanced quadruple in $M'$ can appear at most $\binom{n-3}{n-4}$ times in this process. Hence, the number of balanced quadruples in $M'$ is at least
\begin{equation*}
    \left. \binom{n+1}{n} \cdot H(n) \middle/ \binom{n-3}{n-4} \right. = \frac{1}{4} (m+1)m^2(m-1) = H(n+1).
\end{equation*}

\subsection{Finding an equator}

Let there be $M = (E,\chi)$ with $|E|=n$. We take a representation $\{p_e\}$ and $\{C_{ab}\}$ of $M$ in Lemma~\ref{central}.

\begin{claim}
    There is a centrally symmetric pseudocircle $H \subset \mathbb{S}^2$ that contains no crossing point of the arrangement $\{C_{ab}\}$, whose two sides contain $m+1$ and $m$ points from $\{p_e\}$ respectively, and such that $H \cup \{C_{ab}\}$ is an arrangement of pseudocircles.
\end{claim}

\begin{figure}[t]
\centering
\definecolor{carrierblue}{RGB}{0,157,215}
\definecolor{pushred}{RGB}{237,28,44}
\begin{tikzpicture}[scale=.9,line cap=round,line join=round]
  \coordinate (O) at (3.5,2.3);

  \draw[draw=carrierblue,line width=1.35pt] (O) circle[radius=2cm];

  \draw[draw=pushred,line width=2.7pt,samples=241,smooth,
        variable=\t,domain=0:360]
    plot ({3.5+(2+.34*(6*cos(\t))/sqrt(1+36*cos(\t)*cos(\t)))*cos(\t)},
          {2.3+(2+.34*(6*cos(\t))/sqrt(1+36*cos(\t)*cos(\t)))*sin(\t)})
    --cycle;

  \coordinate (pa) at ({3.5+2*cos(145)},{2.3+2*sin(145)});
  \coordinate (pb) at ({3.5+2*cos(215)},{2.3+2*sin(215)});
  \fill[carrierblue] (pa) circle (3.2pt);
  \fill[carrierblue] (pb) circle (3.2pt);
  \node[carrierblue,font=\large,anchor=east] at
    ({3.5+2.28*cos(145)},{2.3+2.28*sin(145)}) {$p_a$};
  \node[carrierblue,font=\large,anchor=east] at
    ({3.5+2.28*cos(215)},{2.3+2.28*sin(215)}) {$p_b$};
\end{tikzpicture}
\caption{$H$ (red and heavier) is obtained by perturbing $C_{ab}$ (blue and thinner), and has both $p_a$ and $p_b$ on the same side.}
\label{equator}
\end{figure}
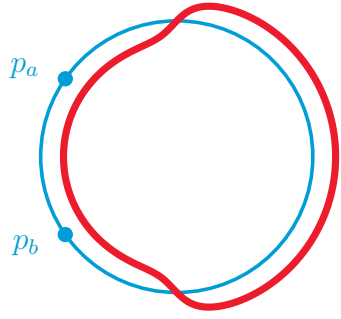

To find such an $H$, fix $a\in E$ and consider, in cyclic order, the $2(n-1)$ local branches at $p_a$ of the curves $C_{ab}$, $b\neq a$. Orient a branch away from $p_a$ and label it by the number of points $p_c$, $c\not\in\{a,b\}$, on its right. The two branches of the same $C_{ab}$ have labels whose sum is $2m-1$. Moreover, consecutive labels differ by at most one. Hence, there must be a particular branch with label $m-1$. Let $C_{ab}$ be the pseudocircle supporting this label. We can slightly perturb $C_{ab}$ to obtain the wanted $H$ such that both $p_a$ and $p_b$ are in the same side of $H$ contains the $m-1$ points labelling the previous particular branch; see Figure~\ref{equator}.

\subsection{Signed affinization}

Let $H^+$ be the side of $H$ containing $m+1$ points from $\{p_e\}$ and $H^-$ be the other side. We give signs to elements in $E$ by defining $\sigma_e = +1$ if $p_e \in H^+$ and $\sigma_e = -1$ if $p_e \in H^-$. We also define $q_e = \sigma_e \cdot p_e \in H^+$ and $L_{ab} = C_{ab} \cap H^+$. We consider a homeomorphism $H^+ \simeq \mathbb{R}^2$, and by abuse of notation, let $q_e$ and $L_{ab}$ also denote their image after this homeomorphism. Because we choose $H$ to be centrally symmetric and generic, $\{q_e\}$ and $\{L_{ab}\}$ form a generalized configuration in $\mathbb{R}^2$.

Now, by Lemma~\ref{allowable}, we have an allowable sequence $\pi_0, \pi_1, \dots, \pi_T$ on $E$. For fixed $\{a,b\} \in \binom{E}{2}$, let $U$ and $V$ be the prefix and suffix of the swap involving them respectively. According to Lemma~\ref{allowable}, $\{q_u\}_{u\in U}$ and $\{q_v\}_{v\in V}$ are separated by $L_{ab}$ in $\mathbb{R}^2$, hence also separated by $C_{ab}$ on $\mathbb{S}^2$. Combining this observation with the last condition of Lemma~\ref{central}, we have a sign $\eta_{ab}$ such that \begin{equation*}
    \chi(a,b,u) = \sigma_u\eta_{ab}, ~\forall~ u \in U; \quad \chi(a,b,v) = -\sigma_v\eta_{ab}, ~\forall~ v\in V.
\end{equation*} We define \begin{equation*}
    Q_{ab} = \sum_{c \in E\setminus\{a,b\}} \chi(a,b,c) \quad\text{and}\quad \Delta_{ab} = 2\cdot \sum_{u\in U} \sigma_u + \sigma_a +\sigma_b -1.
\end{equation*} Using the previous identity, we can compute
\begin{align}
    |Q_{ab}| &= \left| \sum_{u\in U} \sigma_u\eta_{ab} - \sum_{v\in V} \sigma_v\eta_{ab} \right| = \left| \sum_{u\in U} \sigma_u - \sum_{v\in V} \sigma_v \right| \nonumber\\
    &= \left| \sum_{u\in U} \sigma_u - \left( \sum_{e\in E\setminus\{a,b\}} \sigma_e - \sum_{u\in U} \sigma_u \right) \right| \nonumber\\
    &= \left| 2\cdot \sum_{u\in U} \sigma_u - \left( \sum_{e\in E} \sigma_e - \sigma_a - \sigma_b \right) \right| \nonumber\\
    &= |\Delta_{ab}|.\label{quantitybridge}
\end{align}

\subsection{The energy identity}

For $\{a,b\} \in \binom{E}{2}$, let $s_{ab} = 2$ if $\sigma_a = \sigma_b$ and $s_{ab} = -1$ if $\sigma_a \neq \sigma_b$. We define \begin{equation*}
    \mathcal{E} =\sum_{\{a,b\} \in \binom{E}{2}} s_{ab} Q_{ab}^2.
\end{equation*}

\begin{claim}\label{energy} If $B$ is the number of balanced quadruples in $M$, then \begin{equation*}
    \mathcal{E} = 8 \left( B - \frac{1}{4}m^2(m-1)^2 \right) + m(m-1).
\end{equation*}
\end{claim}

To prove this claim, we first expand the squares
\begin{align}
    \mathcal{E} &= \sum_{\{a,b\} \in \binom{E}{2}}s_{ab} \left( \sum_{c \in E\setminus \{a,b\} } \chi(a,b,c) \right)^2 \nonumber \\
    &= \sum_{\{a,b\} \in \binom{E}{2}}s_{ab} \sum_{c \in E\setminus \{a,b\}} (\chi(a,b,c))^2 + 2 \cdot \sum_{\{a,b\} \in \binom{E}{2}}s_{ab} \left(\sum_{\{c,d\} \in \binom{E\setminus \{a,b\}}{2}} \chi(a,b,c) \chi(a,b,d)\right) \nonumber\\
    &= \sum_{\{a,b\} \in \binom{E}{2}}s_{ab} (2m-1) + 2 \cdot \sum_{\{a,b\} \in \binom{E}{2}}s_{ab} \left(\sum_{\{c,d\} \in \binom{E\setminus \{a,b\}}{2}} \chi(a,b,c) \chi(a,b,d)\right) \nonumber\\
    &= m (m-1) (2m-1) + 2 \cdot \sum_{\{a,b\} \in \binom{E}{2}}s_{ab} \left(\sum_{\{c,d\} \in \binom{E\setminus \{a,b\}}{2}} \chi(a,b,c) \chi(a,b,d)\right). \label{energyeq1}
\end{align} We used $|\{e\in E:~ \sigma_e = +1\}| = m+1$ and $|\{e\in E:~ \sigma_e = -1\}| = m$ in above computation.

Next, we compute the contribution $\mathcal{E}_Q$ of a quadruple $Q = \{a,b,c,d\}$ to the second term of \eqref{energyeq1}. With the arbitrary order $a \prec b \prec c \prec d$, we look at its circuit $X_a = \chi(b,c,d)$, $X_b = -\chi(a,c,d)$, $X_c = \chi(a,b,d)$, and $X_d = -\chi(a,b,c)$. Due to the alternating property of $\chi$, we have $X_i X_j = -\chi(k,\ell,i)\chi(k,\ell,j)$ for any $(i,j,k,\ell)$ as a permutation of $(a,b,c,d)$. So we have \begin{equation*}
    \mathcal{E}_Q = -2(s_{ab}X_cX_d + s_{ac}X_bX_d + s_{ad}X_bX_c +s_{bc}X_aX_d + s_{bd}X_aX_c + s_{cd}X_aX_b).
\end{equation*} Using $s_{ab} = (1+3\sigma_a\sigma_b)/2$, we can rewrite \begin{equation}
    \mathcal{E}_Q = -\frac{\prod_{e\in Q} X_e}{2}\left( \left(\sum_{e\in Q} X_e\right)^2 + 3\left( \sum_{e\in Q} \sigma_e X_e \right)^2 - 16 \right). \label{energyeq2}
\end{equation} This means the contribution $\mathcal{E}_Q$ only depends on the types of \begin{equation*}
    X = (X_a,X_b,X_c,X_d) \quad\text{and}\quad \sigma X = (\sigma_a X_a,\sigma_b X_b,\sigma_c X_c,\sigma_d X_d).
\end{equation*} Here, the type of a sequence of signs is defined as the smaller number between the plus-count and the minus-count in it.

We argue that $\sigma X$ cannot be of type 0. Otherwise, we have \begin{equation*}
    \sigma_a X_a \sigma_b X_b = +1 \implies \sigma_b \chi(c,d,b) \sigma_a\chi(c,d,a) = -1,
\end{equation*} which implies, by the last condition of Lemma~\ref{central}, the $L_{cd}$ separates $q_a$ and $q_b$ in $\mathbb{R}^2$. The same can be concluded for all pseudolines induced by $\{q_a,q_b,q_c,q_d\}$, but such a generalized configuration is non-existent. Now, we compute the contribution using \eqref{energyeq2}. Suppose $X$ is of type 0 and $\sigma X$ is of type 1, we have \begin{equation*}
    \prod_{e\in Q} X_e = 1, \quad \left(\sum_{e\in Q} X_e\right)^2 = 16, \quad \left( \sum_{e\in Q} \sigma_e X_e \right)^2=4,
\end{equation*} which implies $\mathcal{E}_Q = -6$. Through the same process, we obtain the values of $E_Q$ for all possible type combinations in the following table.
\begin{equation*}
\begin{array}{|c|c|c|c|}
 \hline &\text{type}(X)=0&\text{type}(X)=1&\text{type}(X)=2\\ \hline
 \text{type}(\sigma X)=1&    -6&0&2\\ \hline
 \text{type}(\sigma X)=2&    0&-6&8\\ \hline
\end{array}
\end{equation*}

Let $b(Q) = 1$ if $Q$ is a balanced quadruple, and $b(Q) = 0$ otherwise. Let $o(Q) = 1$ if the plus-count in $(\sigma_a,\sigma_b, \sigma_c, \sigma_d)$ is odd, and $o(Q) = 0$ otherwise. We summarize the table above into \begin{equation*}
    \mathcal{E}_Q = 8\cdot b(Q) - 6\cdot o(Q).
\end{equation*} Substituting this back to the second term of \eqref{energyeq1}, we get \begin{align*}
    \mathcal{E} &= m (m-1) (2m-1) + \sum_{Q \in \binom{E}{4}} (8\cdot b(Q) - 6\cdot o(Q))\\
    &= m (m-1) (2m-1) + 8\cdot \sum_{Q \in \binom{E}{4}} b(Q) - 6\cdot \sum_{Q \in \binom{E}{4}} o(Q)\\
    &= m (m-1) (2m-1) + 8B - 6 \left( m\binom{m+1}{3} + (m+1)\binom{m}{3} \right)\\
    &= 8 \left( B - \frac{1}{4}m^2(m-1)^2 \right) + m(m-1).
\end{align*}

\subsection{The swap inequality}

Recall that we have an allowable sequence $\pi_0$, $\pi_1$, $\dots$, $\pi_T$ on $E$. For a swap $\pi_{t-1} \to \pi_{t}$, whose pair and prefix written as $\{a,b\}$ and $U$ respectively, we compute its $\Delta$-value by $2\cdot \sum_{u\in U} \sigma_u + \sigma_a + \sigma_b - 1$. For $\{a,b\} \in \binom{E}{2}$, $\Delta_{ab}$ equals the $\Delta$-value of the unique swap involving $a$ and $b$ in the allowable sequence.

\begin{claim}\label{swap}
For a fixed $1\leq k \leq m-1$, among the pairs with $|\Delta_{ab}| \geq 2k+1$, let $S_k$ count the same-sign pairs and $D_k$ count the opposite-sign pairs, then $D_k \leq 2 S_k$.
\end{claim}

First, we extend the allowable sequence by setting $\pi_{T+t}$ as the reverse of $\pi_{t}$ for $1\leq t\leq T$. Since $\pi_T$ is the reverse of $\pi_0$, we have $\pi_{2T} = \pi_0$. Moreover, $\pi_{T+t-1} \to \pi_{T+t}$ is a swap involving the same pair as $\pi_{t-1} \to \pi_t$. Let $\Delta_t$ be the $\Delta$-value of the $t$-th swap, then we have $\Delta_{T+t} = -\Delta_t$. Therefore, among all swaps with $\Delta$-value at least $2k+1$ (absolute value is not applied here), $S_k$ and $D_k$ count the same-sign swaps and opposite-sign swaps respectively. We shall assign each qualifying opposite-sign swap to a qualifying same-sign swap, and make sure each same-sign swap gets at most two assignments, hence proving $D_k \leq 2 S_k$.

We visualize the extended allowable sequence by placing $2T$ points equidistantly on a circle to represent $\pi_0$, $\pi_1$, $\dots$, $\pi_{2T-1}$ clockwise. The gaps between them represent swaps. Fix a qualifying gap $\Gamma$ involving a positive element $a$ and a negative element $b$. This gap and its antipodal gap divide the circle into two semicircles, and on exactly one of them, all orders have $a \prec b$.

For an order $\pi$ in the extended allowable sequence, let $i_a$ be the rank of $a$ among the positive elements, and $j_b$ be the rank of $b$ among the negative elements. The $d$-value of $\pi$ is computed by $d = i_a - j_b$. At the gap $\Gamma$, both orders have the same $d$-value. As the prefix contains $i_a - 1$ positive elements and $j_b - 1$ negative elements, $\Delta$-value of $\Gamma$ is $2d - 1$. Since $\Delta$-value of $\Gamma$ is at least $2k+1$, the $d$-value of the two orders around $\Gamma$ is at least $k+1$. Under reversal, the two ranks become $m+2-i_a$ and $m+1-j_b$ respectively. So the $d$-value of the two orders around the antipodal gap of $\Gamma$ is at most $k$.

Now we look at the semicircle where $a \prec b$. Its two endpoints have $d$-values at most $k$ and at least $k+1$ respectively. We traverse along this semicircle from the endpoint with lower $d$-value to the endpoint with higher $d$-value. We remark that the direction of traversal may be counter-clockwise on the extended allowable sequence. During this process, the $d$-value changes only at gaps representing same-sign swaps involving $a$ or $b$, and each change has magnitude one. Hence, we must encounter a gap when the $d$-value changes from $k$ to $k+1$, we shall argue that the $\Delta$-value of this same-sign gap is at least $2k+1$. We shall assign $\Gamma$ to this target gap. Let $\pi_x$ and $\pi_y$ be the orders around the target gap with $d$-values $k$ and $k+1$ respectively.

Suppose the target swap involves $a$, and it swaps the $i$-th positive element with the $(i+1)$-th positive element. Since $\pi_x$ has smaller $d$-value than $\pi_y$, $a$ must be the $i$-th positive element in $\pi_x$, so $b$ must be the $(i-k)$-th negative element in $\pi_x$. Let $z$ be the number of negative elements preceding $a$ in $\pi_x$. Since we are on the semicircle where $a \prec b$, $z \leq i-k-1$. The $\Delta$-value of the target swap is \begin{equation*}
    2(i-1 - z) +1 +1 -1 \geq  2(i-1 - (i-k-1)) +1 +1 -1 = 2k+1.
\end{equation*} If another opposite-sign swap is assigned to this target swap, it must involve the $(i-k)$-th negative element in $\pi_x$, since $i$ is determined by the target swap itself and $k$ is fixed. It must also involve either the $i$-th or the $(i+1)$-th positive element in $\pi_x$. Hence this target swap will get at most two assignments.

Suppose the target swap involves $b$, and it swaps the $j$-th negative element with the $(j+1)$-th negative element. Since $\pi_y$ has larger $d$-value than $\pi_x$, $b$ must be the $j$-th negative element in $\pi_y$, so $a$ must be the $(j+k+1)$-th positive element in $\pi_y$. Let $z$ be the number of positive elements preceding $b$ in $\pi_y$. Since we are on the semicircle where $a \prec b$, $z \geq j+k+1$. The $\Delta$-value of the target swap is \begin{equation*}
    2(z - (j-1)) -1 -1 -1 \geq  2((j+k+1) - (j-1)) -1 -1 -1 = 2k + 1.
\end{equation*} For a similar reason, this target swap will get at most two assignments.

\subsection{Completing the proof}

Since $Q_{ab}$ is a sum of $2m-1$ signs, it is odd and $|Q_{ab}| \leq 2m-1$. For a positive odd integer $z$, notice the identity \begin{equation*}
    z^2 = 1 + 8\cdot \sum_{k=1}^{\lfloor z/2\rfloor} k.
\end{equation*} Using this identity, \eqref{quantitybridge}, and previous definitions, we can compute \begin{align*}
    \mathcal{E} &= \sum_{\{a,b\} \in \binom{E}{2}} s_{ab}Q_{ab}^2 =\sum_{\{a,b\} \in \binom{E}{2}} s_{ab} \left( 1 + 8\cdot \sum_{k=1}^{\lfloor |Q_{ab}|/2 \rfloor}k \right) \\
    &= \sum_{\{a,b\} \in \binom{E}{2}} s_{ab} + 8\cdot \sum_{\{a,b\} \in \binom{E}{2}} s_{ab} \sum_{k=1}^{\lfloor |Q_{ab}|/2 \rfloor}k\\
    &= m(m-1) + 8\cdot \sum_{\{a,b\} \in \binom{E}{2}} s_{ab} \sum_{k=1}^{\lfloor |\Delta_{ab}|/2 \rfloor}k\\
    &= m(m-1) + 8\cdot \sum_{k=1}^{m-1} k(2S_k - D_k),
\end{align*} where $S_k$ is the number of same-sign pairs and $D_k$ is the number of opposite-sign pairs among the pairs $\{a,b\}$ with $|\Delta_{ab}| \geq 2k+1$.

Now, Claim~\ref{swap} implies $\mathcal{E} \geq m(m-1)$, which implies, by Claim~\ref{energy}, the number of balanced quadruples in $M$ is at least \begin{equation*}
    \frac{1}{4}m^2(m-1)^2 = H(n).
\end{equation*}

\bigskip \noindent {\bf Acknowledgement.} The author used generative AI tools ChatGPT 5.5 and 5.6 for proof exploration and writing betterment.

\end{document}